\documentclass[letterpaper, 10 pt, conference]{ieeeconf}  

\IEEEoverridecommandlockouts                              
\usepackage{mathrsfs}
\usepackage{txfonts}
\usepackage [dvips] {graphicx}

\title{\LARGE \bf
Multi-Sparse Signal Recovery for Compressive Sensing
}

\author{Yipeng Liu$^{1}$, Ivan Gligorijevic$^{1}$, Vladimir  Matic$^{1}$, Maarten De Vos$^{1,~2}$, and Sabine Van Huffel$^{1}$
\thanks{*This work was supported by Research Council KUL: GOA MaNet, CoE EF/05/006 Optimization in Engineering (OPTEC), PFV/10/002 (OPTEC),  IDO 08/013 Autism, several PhD/postdoc and fellow grants; Flemish Government: FWO: PhD/postdoc grants, projects:  FWO G.0302.07 (SVM), G.0341.07 (Data fusion), G.0427.10N (Integrated EEG-fMRI), G.0108.11 (Compressed Sensing) G.0869.12N (Tumor imaging) research communities (ICCoS, ANMMM); IWT: TBM070713-Accelero, TBM070706-IOTA3, TBM080658-MRI (EEG-fMRI), PhD Grants; IBBT; Belgian Federal Science Policy Office: IUAP P6/04 (DYSCO, `Dynamical systems, control and optimization', 2007-2011); ESA AO-PGPF-01,  PRODEX (CardioControl) C4000103224
EU:  RECAP 209G within INTERREG IVB NWE programme, EU HIP Trial FP7-HEALTH/ 2007-2013 (n$бу$ 260777) ( Neuromath (COST-BM0601); BIR$ \& $D Smart Care; Alexander von Humboldt  stipend.
}
\thanks{$ ^1 $All the authors are with KU Leuven, Dept. of Electrical Engineering (ESAT) SCD-SISTA and IBBT Future Health Department, Kasteelpark Arenberg 10, box 2446, 3001 Leuven, Belgium. }%
\thanks{$ ^2 $MDV is also with Neuropsychology, Dept. of Psychology, University of Oldenburg, Oldenburg, Germany.}
}

\begin{document}

\maketitle
\thispagestyle{empty}
\pagestyle{empty}

\begin{abstract}
Signal recovery is one of the key techniques of Compressive sensing (CS). It reconstructs the original signal from the linear sub-Nyquist measurements. Classical methods exploit the sparsity in one domain to formulate the L0 norm optimization. Recent investigation shows that some signals are sparse in multiple domains. To further improve the signal reconstruction performance, we can exploit this multi-sparsity to generate a new convex programming model. The latter is formulated with multiple sparsity constraints in multiple domains and the linear measurement fitting constraint. It improves signal recovery performance by additional \emph{a priori} information. Since some EMG signals exhibit sparsity both in time and frequency domains, we take them as example in numerical experiments. Results show that the newly proposed method achieves better performance for multi-sparse signals.

\end{abstract}

\section{INTRODUCTION}

Compressive sensing (CS) has attracted considerable attention in signal processing. It employs nonadaptive linear projections that preserve the structure of the signal; the signal is then reconstructed from these projections using nonlinear methods. Rather than first sampling at a high Nyquist rate and then compressing the sampled data, it directly senses the data in a compressed form with a lower sampling rate. CS provides a new promising framework for acquiring signals. Signal recovery, as one of the key techniques of CS, reconstructs the original signal from the linear sub-Nyquist measurements \cite{candes_intro_cs}.

\begin{figure}[!h]
 \centering
 \includegraphics[scale = 0.40]{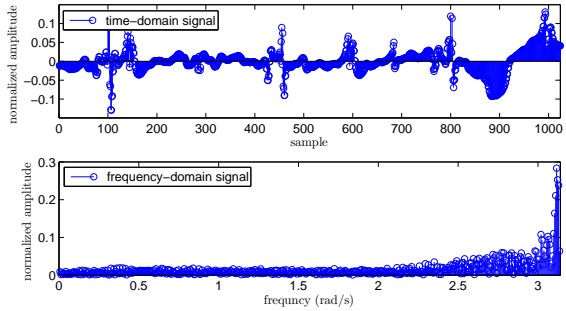}
 \caption{ An example of EMG data from a healthy person: $ EMG - healthy $.}
 \label{figure1}
\end{figure}

In classical signal recovery, sparsity is exploited by formulating an L1-norm optimization problem. Only one sparsity constraint is used with a linear measurement fitting constraint \cite{marvasti_unified_ssp}. But some signals are sparse in more than one domain. For example, some electromyography (EMG) signals are sparse in both time and frequency domains \cite{kimura_emg} \cite{zwarts_emg} \cite{salman_cs_emg}, as shown in Fig. \ref{figure1}; some microwave signals are sparse in both frequency and space domains \cite{liu_mn_bf} \cite{liu_stv_bf}. Taking into consideration the multi-sparsity, we use multiple L1 norm minimization based sparsity constraints to encourage sparse distribution in all the corresponding domains. As more \emph{a priori} information is used, the recovery performance would be enhanced. Numerical experiments demonstrate the proposed method has a better performance than previous methods.
%

\section{COSPARSE ANALYSIS SIGNAL MODEL}

Sparsity exists in many signals. It means that many of the representation coefficients are close to or equal to zero, when the signal is represented in some domain. Traditionally, a synthesis representation model decomposes the signal into a linear combination of a few columns chosen from a predefined dictionary (representation matrix). Recently, a new signal model, called cosparse analysis model, was proposed in \cite{nam_analysis_model}. In this new sparse representation, an analysis operator multiplying the measurements leads to a sparse outcome. Let signal in discrete form be expressed as:

\begin{equation}\label{2_1_sparse representation}
{\bf{\theta }} = {\bf{\Psi x}}
\end{equation}
where $ {\bf{x}} \in {\texttt{R}^{N \times 1}} $  is the original signal obtained at Nyquist sampling rate; $ {\bf{\Psi }} \in {\texttt{C}^{L \times N}} $ is the analysis operator; $ {\bf{\theta }} \in {\texttt{C}^{L \times 1}} $ is the resulting sparse representative vector, i.e. most of the elements of $ {\bf{\theta }} $ are almost zero. Here $ L \ge N\ $. In a practical CS system, the analogue baseband signal \emph{x}(\emph{t}) is sampled using an analogue-to-information converter (AIC) \cite{laska_aic}. The AIC can be conceptually modeled as an analogue-to-digital converter (ADC) operating at Nyquist rate, followed by a CS operation. Then the random sub-Nyquist measurement vector $ {\bf{y}} \in {\texttt{R}^{M \times 1}} $ is obtained directly from the continuous-time signal \emph{x}(\emph{t}) by AIC. For demonstration convenience, we formulate the sampling as:
\begin{equation}\label{2_2_random sampling}
{\bf{y}} = {\bf{\Phi x}}
\end{equation}
where $  {\bf{\Phi }} \in {\texttt{R}^{M \times N}} $ is the measurement matrix; $ M \ll N\ $. Three frequently used examples are: Gaussian matrix; Bernoulli matrix and partial Fourier matrix.

\section{MULTI-SPARSE SIGNAL RECOVERY}

After obtaining the random samples from AIC, they are sent to the digital signal processor (DSP) to get the signal. The classical sparse signal recovery model can be formulated as:

\begin{equation}\label{3_1_L0}
\begin{array}{c}
\mathop {\min }\limits_{\bf{x}} {\left\| {{\bf{\Psi x}}} \right\|_0}\\
{\rm{s}}~{\rm{. t}}~{\rm{.  }}~~{\bf{y}} = {\bf{\Phi x}}
\end{array}
\end{equation}
where the L0 norm $ {\left\| {\bf{\theta }} \right\|_0} $, counting the number of nonzero entries of the vector $ {\bf{\theta }} = {\left[ {{\theta _1},{\theta _2}, \cdots ,{\theta _N}} \right]^T} $, encourages sparse distribution. However, (\ref{3_1_L0}) is NP-hard unfortunately.

 Mainly three groups of algorithms exist to solve (\ref{3_1_L0}) \cite{liu_phd}. The first one is convex programming, such as basis pursuit (BP), Dantzig selector (DS); the second one constitutes of greedy algorithms, such as matching pursuit (MP), orthogonal matching pursuit (OMP); the third one includes hybrid methods, such as CoSaMP, stage-wise OMP (StOMP). In these algorithms, convex programming has the best reconstruction accuracy; greedy algorithms have the least computational complexity; hybrid methods try to balance the reconstruction accuracy and computational complexity.

\subsection{L1 optimization}

In order to get the highest recovery accuracy, we choose convex programming to do CS. Basis pursuit denoising (BPDN) is the most popular one. It can be formulated as

\begin{equation}\label{3_2_BPDN}
\begin{array}{c}
\mathop {\min }\limits_{\bf{x}} {\left\| {{\bf{\Psi x}}} \right\|_1}\\
{\rm{s}}{\rm{.~t}}{\rm{.~}}~~{\left\| {{\rm{ }}{\bf{y}} - {\bf{\Phi x}}} \right\|_2} \le \varepsilon
\end{array}
\end{equation}
where $  {\left\| {\bf{\theta }} \right\|_1} = \sum\nolimits_m | {{\theta _m}}| $  is the L1 norm of the vector $ {\bf{\theta }} = {\left[ {{\theta _1},{\theta _2}, \cdots ,{\theta _N}} \right]^T} $; $ \varepsilon $ is a nonnegative scalar bounding the amount of noise in the data.

If the signal is sparse in the time domain, we can choose identity matrix as the analysis operator, i.e. $ {\bf{\Psi }} = {\bf{I}} $. Thus, the standard BPDN can be reformulated as:
\begin{equation}\label{3_3_T-L1}
\begin{array}{c}
\mathop {\min }\limits_{\bf{x}} {\left\| {\bf{x}} \right\|_1}\\
{\rm{s}}{\rm{.~t}}{\rm{. }}~~{\left\| {{\rm{ }}{\bf{y}} - {\bf{\Phi x}}} \right\|_2} \le \varepsilon
\end{array}
\end{equation}
Here we call (\ref{3_3_T-L1}) T-L1 optimization. This convex optimization model can be reformulated as:
\begin{equation}\label{3_4_T-L1-sdp}
\begin{array}{c}
\mathop {\min }\limits_{{\bf{x}},{\bf{t}}} {{\bf{1}}^T}{\bf{t}}\\
{\rm{s}}{\rm{.~t}}{\rm{.~~}}{\left\| {{\rm{ }}{\bf{y}} - {\bf{\Phi x}}} \right\|_2} \le \varepsilon \\
 - {\bf{t}} \prec {\bf{x}} \prec {\bf{t}}
\end{array}
\end{equation}
where \textbf{1} is an \emph{N}-by-\emph{1} vector with all elements being 1. (\ref{3_4_T-L1-sdp}) is a semidefinite programming (SDP) problem. It can be solved by software, such as SeDuMi \cite{sturm_sedumi}, cvx \cite{boyd_cvx}, etc.

Similarly if the signal is sparse in the frequency domain, we can also recover it by:
\begin{equation}\label{3_5_F-L1}
\begin{array}{c}
\mathop {\min }\limits_{\bf{x}} {\left\| {{\bf{Fx}}} \right\|_1}\\
{\rm{s}}{\rm{.~t}}{\rm{.~~}}{\left\| {{\rm{ }}{\bf{y}} - {\bf{\Phi x}}} \right\|_2} \le \varepsilon
\end{array}
\end{equation}
where \textbf{F} is the \emph{N}-by-\emph{N} Fourier transform matrix. To distinguish (\ref{3_5_F-L1}) from (\ref{3_3_T-L1}), (\ref{3_5_F-L1}) is named F-L1 optimization. It can be reformulated as an SDP:
\begin{equation}\label{3_6_F-L1-sdp}
\begin{array}{c}
\mathop {\min }\limits_{{\bf{x}},{\bf{t}}} {{\bf{1}}^T}{\bf{Ft}}\\
{\rm{s}}{\rm{.~t}}{\rm{.~~}}{\left\| {{\rm{ }}{\bf{y}} - {\bf{\Phi x}}} \right\|_2} \le \varepsilon \\
 - {\bf{t}} \prec {\bf{Fx}} \prec {\bf{t}}
\end{array}
\end{equation}

\subsection{Multi-L1 optimization}
 To further enhance the performance of signal reconstruction, we can exploit the unique property that some signals are sparse in multiple domains. This \emph{a priori} information may be helpful to improve the signal recovery performance. Here we propose a new optimization model for multi-sparse signal recovery as:
\begin{equation}\label{3_7_Multi-L0}
\begin{array}{c}
\mathop {\min }\limits_{\bf{x}} \sum\limits_{p = 1}^P {\left( {{\lambda _p}{{\left\| {{{\bf{\Psi }}_p}{\bf{x}}} \right\|}_0}} \right)} \\
{\rm{s}}{\rm{.~t}}{\rm{.~~}}{\left\| {{\rm{ }}{\bf{y}} - {\bf{\Phi x}}} \right\|_2} \le \varepsilon
\end{array}
\end{equation}
where \emph{P} is the number of analysis operators which generate sparse outcome; $ {\lambda _p} $ , \emph{p} = 1, 2, ... , \emph{P}, is the parameter balancing the different sparsity constraints. Here we call (\ref{3_7_Multi-L0}) multi-L0 optimization. Since more \emph{a priori} information is used, we expect to achieve better reconstruction performance.

Transforming the nonconvex multi-L0 optimization (\ref{3_7_Multi-L0}) into a convex programming one, we get
\begin{equation}\label{3_78_Multi-L1}
\begin{array}{c}
\mathop {\min }\limits_{\bf{x}} \sum\limits_{p = 1}^P {\left( {{\lambda _p}{{\left\| {{{\bf{\Psi }}_p}{\bf{x}}} \right\|}_1}} \right)} \\
{\rm{s}}{\rm{. t}}{\rm{.~~}}{\left\| {{\rm{ }}{\bf{y}} - {\bf{\Phi x}}} \right\|_2} \le \varepsilon
\end{array}
\end{equation}
We call (\ref{3_78_Multi-L1}) multi-L1 optimization. It can be rewritten as an SDP:

\begin{equation}\label{3_78_Multi-L1-SDP}
\begin{array}{c}
\mathop {\min }\limits_{{\bf{x}},{{\bf{t}}_1},...,{{\bf{t}}_P}} \left( {\sum\limits_{p = 1}^P {{\lambda _p}{{\bf{1}}^T}{{\bf{t}}_p}} } \right)\\
{\rm{s}}{\rm{. t}}{\rm{. }}{\left\| {{\rm{ }}{\bf{y}} - {\bf{\Phi x}}} \right\|_2} \le \varepsilon \\
\begin{array}{*{20}{c}}
{ - {{\bf{t}}_1} \prec {{\bf{\Psi }}_1}{\bf{x}} \prec {{\bf{t}}_1}}\\
 \vdots \\
{ - {{\bf{t}}_p} \prec {{\bf{\Psi }}_p}{\bf{x}} \prec {{\bf{t}}_p}}
\end{array}
\end{array}
\end{equation}

\subsection{L1-L1 optimization for EMG signal recovery}

EMG aims at recording of the electrical activity produced by muscles. It is very useful for detection of various pathologies \cite{zwarts_emg}. Long-term EMG monitoring using multiple channels usually requires a very large amount of data for sampling, transmitting, storage and processing. However, the wireless portable recording devices are constrained to have low battery power, small size and limited transmitting power due to the portability requirement and safety constraints. Therefore, real-time data compression is important \cite{bachman_wsbn}.

Some EMG signals are sparse in both time and frequency domains, and CS has been introduced to EMG bio-signals \cite{salman_cs_emg}. But it mainly investigates the effects of the quantization of the random coefficients of the measurement matrix.

Here we apply the multi-L1 optimization to the EMG signal recovery, we set \emph{P} = 2, $ {{\bf{\Psi }}_1} = {\bf{I}} $, and  $ {{\bf{\Psi }}_2} = {\bf{F}} $, the multi-L1 optimization (\ref{3_78_Multi-L1}) reduces to:
\begin{equation}\label{3_8_L1-L1}
\begin{array}{c}
\mathop {\min }\limits_{\bf{x}} \left( {{{\left\| {\bf{x}} \right\|}_1} + {\lambda _2}{{\left\| {{\bf{Fx}}} \right\|}_1}} \right)\\
{\rm{s}}{\rm{.~t}}{\rm{.~~}}{\left\| {{\rm{ }}{\bf{y}} - {\bf{\Phi x}}} \right\|_2} \le {\varepsilon}
\end{array}\
\end{equation}
where $ {\lambda _2} $ is a nonnegative scalar balancing the two L1 norm minimization based sparsity constraints. Here $ {\lambda _2} $ is related to the length of signal \emph{N}. (\ref{3_8_L1-L1}) is called L1-L1 optimization. To solve it, we can reformulate it as:
\begin{equation}\label{3_9_L1-L1-sdp}
\begin{array}{c}
\mathop {\min }\limits_{{\bf{x}},{\bf{t}},{\bf{r}}} \left( {{{\bf{1}}^T}{\bf{t}} + {\lambda _2}{{\bf{1}}^T}{\bf{r}}} \right)\\
{\rm{s}}{\rm{.~t}}{\rm{.~~}}{\left\| {{\rm{ }}{\bf{y}} - {\bf{\Phi x}}} \right\|_2} \le {\varepsilon}\\
 - {\bf{t}} \prec {\bf{x}} \prec {\bf{t}}\\
 - {\bf{r}} \prec {\bf{Fx}} \prec {\bf{r}}
\end{array}
\end{equation}
(\ref{3_9_L1-L1-sdp}) is an SDP. It can be solved by convex optimization software too \cite{sturm_sedumi} \cite{boyd_cvx}.

\section{NUMERICAL EXPERIMENTS}

In the numerical experiments, we use the proposed multi-L1 optimization, the L1 norm optimization and the relaxed least squares (LS) method with $ {\min _{\bf{x}}}{\left\| {\bf{x}} \right\|_2},~{\rm{ s}}{\rm{.~t}}{\rm{.~}}\left\| {{\bf{y}} - {\bf{\Phi x}}} \right\| \le \varepsilon \ $ to recovery a group of multi-sparse signals. Then we compare the signal recovery performance.

The multi-sparse signals are chosen to be the EMG signals which are obtained from the \emph{Physiobank} database \cite{goldberger_phsiobank}.  In \cite{salman_cs_emg}, the static thresholding algorithm is used to reconstruct the EMG signals, whose accuracy is obviously worse than convex relaxation. The measurement matrix  $ {\bf{\Phi }} $ is formed by sampling the i.i.d. entries from a white Gaussian distribution. Here four signal recovery methods, Least Squares (LS) methods with ${\bf{\hat x}} = \mathop {\arg \min }\nolimits_{\bf{x}} {\left\| {\bf{x}} \right\|_2},~{\rm{s}}{\rm{.~t}}{\rm{.~~}}{\left\| {{\bf{y}} - {\bf{\Phi x}}} \right\|_2} $ , T-L1 optimization (\ref{3_3_T-L1}), F-L1 optimization (\ref{3_5_F-L1}), and the newly proposed L1-L1 optimization (\ref{3_8_L1-L1}), are used to reconstruct the EMG signals. $ {\lambda _2} $ is chosen to be 0.05 in order to balance; $ \varepsilon $ is chosen to be $ 5\% $ of the measurement power, i. e. $ \varepsilon  = 0.05{\left\| {\bf{y}} \right\|_2} $.

To quantify the performance of signal recovery, the root mean squared error (RMSE) is calculated via the formula:

\begin{equation}\label{4_1_rmse}
e = \frac{1}{L}\sum\limits_{l = 1}^L {\frac{{{{\left\| {{{\bf{x}}_l} - {{{\bf{\hat x}}}_l}} \right\|}_2}}}{{{{\left\| {{{\bf{x}}_l}} \right\|}_2}}}}
\end{equation}
Here $ {{\bf{x}}_l} $ is the normalized original EMG signal in the \emph{l}-th Monte Carlo simulation; ${{\bf{\hat x}}_l} $ is the normalized estimated EMG signal in the \emph{l}-th Monte Carlo simulation; \emph{L} is the number of Monte Carlo simulations. Because the amount of available data is limited, \emph{L} is chosen to be 40.

\begin{figure}[!h]
 \centering
 \includegraphics[scale = 0.40]{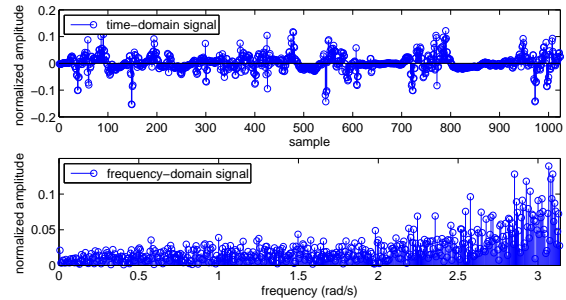}
 \caption{ An example of EMG data from a patient with myopathy: $ EMG - myopathy $.}
 \label{figure2}
\end{figure}

\begin{figure}[!h]
 \centering
 \includegraphics[scale = 0.40]{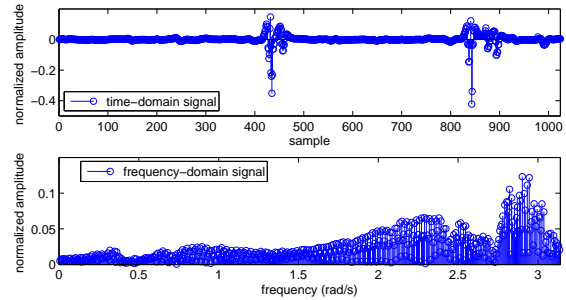}
 \caption{ An example of EMG data from a patient with neuropathy: $ EMG - neuropathy $.}
 \label{figure3}
\end{figure}

Fig. \ref{figure1}, Fig. \ref{figure2} and Fig. \ref{figure3} show three sections of EMG signals of a healthy person ($ EMG - healthy $), a patient with myopathy ($ EMG - myopathy $) and a patient with neuropathy ($ EMG - neuropathy $), respectively. We can see that all three signals are sparse in the time domain. In the frequency domain, $ EMG - healthy $ and $  EMG - myopathy $ signals are sparse but the $ EMG - neuropathy $ signal is not.

\begin{figure}[!h]
 \centering
 \includegraphics[scale = 0.40]{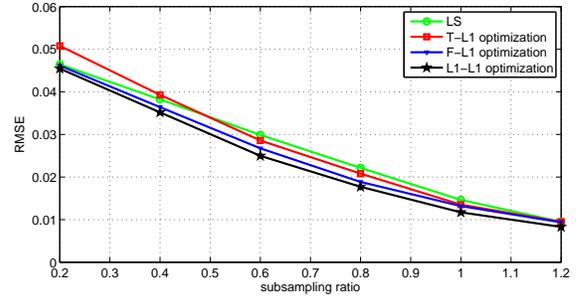}
 \caption{ Signal recovery performance for the data $ EMG - healthy $.}
 \label{figure4}
\end{figure}

\begin{figure}[!h]
 \centering
 \includegraphics[scale = 0.40]{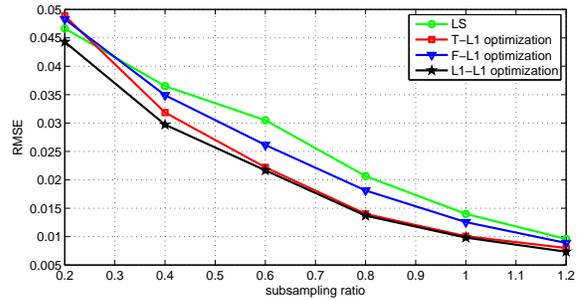}
 \caption{ Signal recovery performance for the data $ EMG - myopathy $.}
 \label{figure5}
\end{figure}

\begin{figure}[!h]
 \centering
 \includegraphics[scale = 0.40]{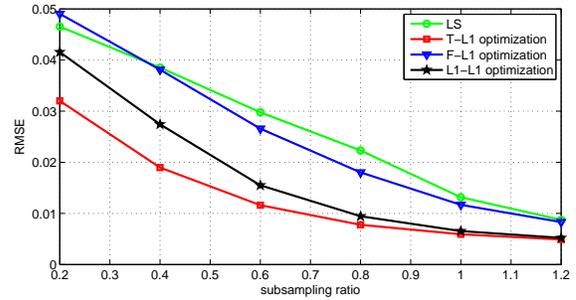}
 \caption{ Signal recovery performance for the data $ EMG - neuropathy $.}
 \label{figure6}
\end{figure}

\begin{figure}[!h]
 \centering
 \includegraphics[scale = 0.40]{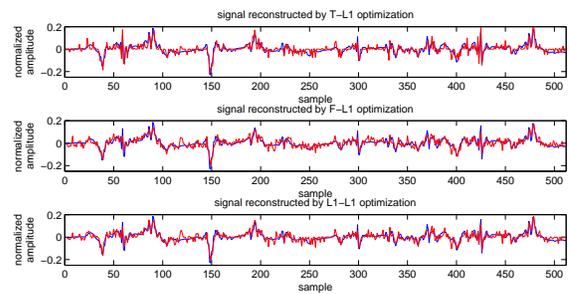}
 \caption{An example of $ EMG - neuropathy $ signal recovery with the sub-sampling ratio = 0.50.}
 \label{figure7}
\end{figure}

Fig. \ref{figure4}, Fig. \ref{figure5} and Fig. \ref{figure6} show the recovery performance of the three different EMG signals. Here the length of the original EMG signal sections is equal to \emph{N} = 512. All RMSE values decrease with the increase of sub-sampling ratio \emph{M/N}. When the sub-sampling ratio reaches 1, it still can not achieve the perfect reconstruction with RMSE = 0, which results from the relaxation of the constraint from $ {\bf{y}} = {\bf{\Phi x}} $ to $ {\left\| {{\bf{y}} - {\bf{\Phi x}}} \right\|_2} \le \varepsilon $. It may be the price for robustness. Besides, because all the EMG data are noisy, and the noiseless signal can not be available in (\ref{4_1_rmse}), the performance may be better than RMSE demonstrates. To present the recovery performance more directly, Fig. \ref{figure7} gives an example of the reconstruction of a section of $  EMG - myopathy $ signal with sub-sampling ratio equals to 0.50. We can see the profile of the signal can be well reconstructed.

In Fig. \ref{figure4}, T-L1 optimization performs better than F-L1 optimization; but in Fig. \ref{figure5}, F-L1 optimization is better than T-L1 optimization. However, L1-L1 optimization is the best of all in both Fig. \ref{figure4} and Fig. \ref{figure5}. In Fig. \ref{figure6}, we can see that L1-L1 optimization is better than F-L1 optimization, but worse than T-L1 optimization. The reason is that the EMG signal here is not sparse in the frequency domain, which can be seen in Fig. \ref{figure3}.

In summary, if the EMG signal is sparse in both time and frequency domains, L1-L1 optimization is the best candidate for compressive EMG signal recovery. Moreover, if the signal is likely to be sparse in multiple domains with a certain degree of uncertainty, the L1-L1 optimization is also a robust choice, because it can at least avoid the worst performance.

In addition, when \emph{M} = 256, the average computing time for T-L1 optimization, F-L1 optimization and the L1-L1 optimization is respectively 2.7116 seconds, 17.2069 seconds and 10.3265 seconds. The computing time of L1-L1 optimization is longer than T-L1 optimization but shorter than F-L1 optimization.

\section{CONCLUSION}

In this paper, we propose a signal recovery method for multi-sparse signals. The newly proposed multi-L1 optimization encourages sparse distribution in multiple domains. Since more\emph{ a priori} information is exploited, the signal recovery performance would be enhanced. Numerical experiments take EMG signals as examples to demonstrate the performance improvement.

In the future, we will analyze the theoretical conditions for successful recovery by the proposed method. Furthermore, we will develop a hybrid method for multi-sparse signal recovery to decrease the computation complexity. The Split Bregman method will be used to accelerate the solution of multi-sparse signal recovery problem.

%
%
%

%
%
%
%

\end{document}